\documentclass[12pt,preprint]{aastex}

\usepackage{amssymb}
\usepackage{graphicx}

\shorttitle{The Metallicity and Colors of M31 GCs}

\shortauthors{Fan et al.}

\begin{document}

\title{The Correlations between the Intrinsic Colors
and Spectroscopic Metallicities of M31 Globular Clusters}

\author{Zhou Fan\altaffilmark{1,2},
  Jun Ma\altaffilmark{1,2},
  Xu Zhou\altaffilmark{1,2} and Zhaoji Jiang\altaffilmark{1,2}}

\altaffiltext{1}{National Astronomical Observatories, Chinese Academy of
  Sciences, A20 Datun Road, Chaoyang District, Beijing 100012, China}

\altaffiltext{2}{Key Laboratory of Optical Astronomy, National
Astronomical Observatories, Chinese Academy of Sciences}

\email{zfan@nao.cas.cn}

\begin{abstract}
  We present the correlations between the spectroscopic
  metallicities and ninety-three different intrinsic colors of M31 globular
  clusters, including seventy-eight BATC colors and fifteen SDSS and near
  infrared $ugriz$K colors. The BATC colors were derived from the archival
  images of thirteen filters (from $c$ to $p$), which were taken by
  Beijing-Arizona-Taiwan-Connecticut (BATC) Multicolor Sky Survey
  with a 60/90 cm f/3 Schmidt telescope. The spectroscopic
  metallicities adopted in our work were from literature. We fitted the
  correlations of seventy-eight different BATC colors and the metallicities
  for 123 old confirmed globular clusters, and the result implies that
  correlation coefficients of twenty-three colors $r>0.7$. Especially,
  for the colors $(f-k)_0$, $(f-o)_0$, and $(h-k)_0$, the correlation coefficients
  are $r>0.8$. Meanwhile, we also note that the correlation coefficients ($r$)
  approach zero for $(g-h)_0$, $(k-m)_0$, $(k-n)_0$, and $(m-n)_0$, which are
  likely to be independent of metallicity. Similarity, we fitted the
  correlations of metallicity and $ugriz$K colors for 127 old confirmed
  GCs. The result indicates that all these colors are metal-sensitive
  ($r>0.7$), of which $(u-$K$)_0$ is the most metal-sensitive color. Our work
  provides an easy way to simply estimate the metallicity from colors.
\end{abstract}

\section{Introduction}
\label{sect:intr}

Globular clusters (hereafter GCs) are the oldest bound stellar
systems in the Milky Way and other galaxies. They provide a fossil
record of the earliest stages of galaxy formation and evolution.
Located at a distance of $\sim780$ kpc \citep{sg98, mac01} away,
Andromeda (M31) is the nearest large spiral galaxy in our local
group. According to the latest RBC v.4.0
\citep{gall04,gall06,gall07}, M31 contains more than one thousand
GCs and candidates. Therefore M31 GCs provide us an ideal laboratory
for us to study the nature of extragalactic GCs and to better
understand the formation and evolution of M31. Furthermore, since
M31 is Sb-type, which is similar to our Galaxy, it may offer us the
clues to formation and evolution history of our Galaxy.

The correlation of metallicity and colors of globular clusters has
been previously studied by many authors: \citet{bh90} fitted the
correlations of metallicities and ($V-$K), ($J-$K) for Galactic GCs
and applied the calibrated relation to M31 GCs to derive the
metallicities of M31 GCs. \citet{barmby00} analyzed the correlations
of ten intrinsic colors ($(B-V)_0$, $(B-R)_0$, $(B-I)_0$, $(U-B)_0$,
$(U-V)_0$, $(U-R)_0$, $(V-R)_0$, $(V-I)_0$, $(J-$K$)_0$,
$(U-$K$)_0$) and metallicity of 88 Galactic GCs. They find that the
correlation coefficients \emph{r} for all these ten colors range
from 0.91 to 0.77, of which $(U-R)_0$ is the best metallicity
indicator while $(V-R)_0$ is not so good. The authors also studied
the $(J-$K$)_0$ and $(V-$K$)_0$ color-metallicity  correlations for
M31 and Milky Way globular clusters. In this work, we attempt to
find the most metal-sensitive colors, which could be used for the
derivation of the metallicities of GCs in the future work. It is
well known that, spectroscopy is relatively expensive in terms of
observational time and is rarely available for many extragalactic
globular clusters. Hence it is useful to determine metallicities
from photometry.

The studies on identification, classification and analysis of the
M31 GCs have been started since the pioneering work of
\citet{hub32}, \citep[see
e.g.][]{vet62,sar77,batt80,batt87,batt93,cra85,barmby00}. These
studies provided a large amount of photometric data in different
photometric systems, i.e. CCD photometry, photoelectric photometry,
and photographic plates, even visual photometry. In order to obtain
homogeneous photometric catalogue of M31 GCs, \citet{gall04, gall06,
gall07} took the photometry of \citet{barmby00} as reference and
transformed others to this reference and compiled the famous The
Revised Bologna Catalogue of M31 GCs and GC candidates (The latest
version is RBC v.4.0), where 654 confirmed GCs and 619 GC candidates
of M31 have been included, and 772 former GC candidates have been
proved to be stars, asterism, galaxy, region HII or extended
cluster. The catalogue also includes the newly discovered star
clusters from \citet{mac06}, \citet{kim07} and \citet{hux08}.
\citet{mac06} reported the discovery of eight remote GCs in the
outer halo of M31 by using the deep ACS images; \citet{kim07} found
1164 GCs and GC candidates in M31 with KPN 0.9 m telescope and the
WIYN 3.5 m telescope, of which 559 are previously known GCs and 605
are newly found GC candidates; later, \citet{hux08} detected 40 new
GCs in the halo of M31 with Isaac Newton Telescope and
Canada-France-Hawaii Telescope data. Recently, \citet{cald09}
presented a new catalogue of 670 likely star clusters, stars,
possible stars and galaxies in the field of M31, all with updated
high-quality coordinates being accurate to $0.2''$ based on the
images from the Local Group Survey \citep{massey06} or Digitized Sky
Survey (DSS). Very recently, \citet{peac10} identified M31 GCs with
images from Wide Field CAMera (WFCAM) on the UK Infrared Telescope
and SDSS archives, and performed the photometry for them in SDSS
$ugriz$ and infrared $K$ bands. Furthermore, the authors combined
all the identifications and photometry of M31 GCs from references
and those of their new work and updated the M31 star cluster
catalog, including 416 old confirmed clusters, 156 young clusters
and 373 candidate clusters. Our work is based on the 416 old
confirmed M31 GCs, from which we will select our sample GCs.

Studies of M31 GCs based on BATC observations are also numerous:
\citet{jiang03} presented the BATC photometry of 172 GCs in the
central $\sim 1$ deg$^2$ region of M31 and estimated the ages for
them with the Simple Stellar Population (SSP) models. After that,
\citet{fan09} added six more new BATC observations fields
surrounding the central region and performed the photometry for
thirty GCs, which did not have the broadband photometry before. With
the photometry, the authors, for the first time, suggested the blue
tilt of M31 GCs. \citet{ma09} fitted the ages of thirty-five GCs of
the central M31 field which were not included in \citet{jiang03}
with BATC, 2MASS and GALEX data and the SSP models. Later,
\citet{wang10} performed the photometry for another 104 GCs of M31
and estimated the ages by fitting the SEDs with SSP models,
revealing the existence of young, the intermediate-age and the old
populations in M31.

In this paper, first we used the BATC data to analyze the
correlations of spectroscopic metallicities and the colors of M31
GCs. Moreover, we did the similar work with the $ugriz$K band data.
The paper is organized as follows: Sect. \ref{sect:data} describes
the data utilized in our work, including the spectroscopic
metallicity from literature, BATC archival images and the data
reduction, the $ugriz$K photometry from literature, along with the
reddening of M31 GCs. Sect. \ref{sect:ana} present the analysis and
the results on the correlations of the metallicities and intrinsic
colors for M31 GCs. The summary and remarkable conclusions of our
work are in Sect. \ref{sect:sum}.

\section{Data}
\label{sect:data}

\subsection{The Spectroscopic Metallicities Adopted}

The spectroscopic metallicities used in our analysis were from
\citet{per02}, who measured the metallicites for over 200 GCs in M31
with William Herschel 4.2 m telescope in La Palma, Canary Islands
from Nov. 3 to 6, 1996. The Wide Field Fibre Optic Spectrograph
(WYFFOS) and two gratings (the H2400B 2400 line grating and the
R1200R 1200 line grating) were applied for the observations. The
dispersion of H2400B 2400 line grating is 0.8 {\AA} pixel$^{-1}$ and
spectral resolution is 2.5 {\AA} over the range $3700-4500$ {\AA}.
The dispersion of the R1200R 1200 line grating observations is 1.5
{\AA} pixel$^{-1}$ and the resolution is 5.1 {\AA} over the spectral
range $4400-5600$ {\AA}. The total wavelength coverage of the
spectra is $\sim3700-5600$ {\AA}, and it can coverage the spectral
features such as CN, H$\beta$, the Mg $b$ triplet, iron lines. For
homogenous reason, we only adopted the spectroscopic metallicities
from \citet{per02} in our work.

\subsection{The BATC Archival Images and Photometry}
\label{sect:batc}

The 15-band archival images of M31 field were obtained by BATC
60/90cm f/3 Schmidt telescope at Xinglong site of Hebei Provence,
China, during 1995 February $-$ 2008 March. The total exposure time
of archival M31 BATC images is 143.9 hours with 447 frames and
covering $\sim6~\rm deg^2$ sky field. The observations and dataset
are described in detail by \citet{fan09}. The fifteen
intermediate-band filters of BATC system are specifically designed
to avoid most bright night-sky emissions, covering the wavelength
range of $\sim$3,000 to $\sim$10,000 {\AA} \citep[see e.g.
][]{fan96,yan00,zhou03,fan09,ma09}.

Before this work, \citet{fan09} have performed usual data reduction
for these images, including bias subtraction and flat fielding of
the CCD images, with an automatic data-reduction routine PIPELINE~I,
developed for the BATC sky survey based on IRAF. The BATC magnitudes
are well defined and obtained in a similar way as for the
spectrophotometric AB magnitude system \citep[for details see
][]{yan00,zhou03,ma09}. For BATC fifteen intermediate-band filters
$abcdefghighmnop$ of the central field of M31 (M31-1 field),
\citet{fan09} calibrated the combined images with \citet{og83}
standard stars \citep[see e.g. ][]{fan96,yan00,zhou03}, while for
the M31-2 to M31-7 fields, the secondary standard stars of the
overlapping field were used for the flux calibration \citep[for
details see ][]{fan09}.

In this work, we performed the aperture photometry for M31 GCs with
the BATC photometric routine PIPELINE~II, which is the revised
version of IRAF \emph{daofind} and \emph{daophot} for BATC system
data reduction. For aperture photometry parameters, we adopted the
aperture radius $r=5''.1$ and the inner sky radius and outer sky
radius are $\sim13''.6$ and $\sim22''.1$, respectively, which are
the same parameters as those used in
\citet{jiang03,fan09,ma09,wang10} for consistency. After careful
checking all the images, we found the signal-to-noise ratios (SNR)
of BATC $a$-band and $b$-band images are not high enough for our
study, thus we do not included these images in the following work.
Then, we obtained the BATC magnitudes of our sample GCs in the
thirteen bands (from $c$ to $p$), covering the wavelength from $\sim
4,000$ to $\sim 10,000$ {\AA}.

\begin{figure}
\center
\includegraphics[scale=.7,angle=-90]{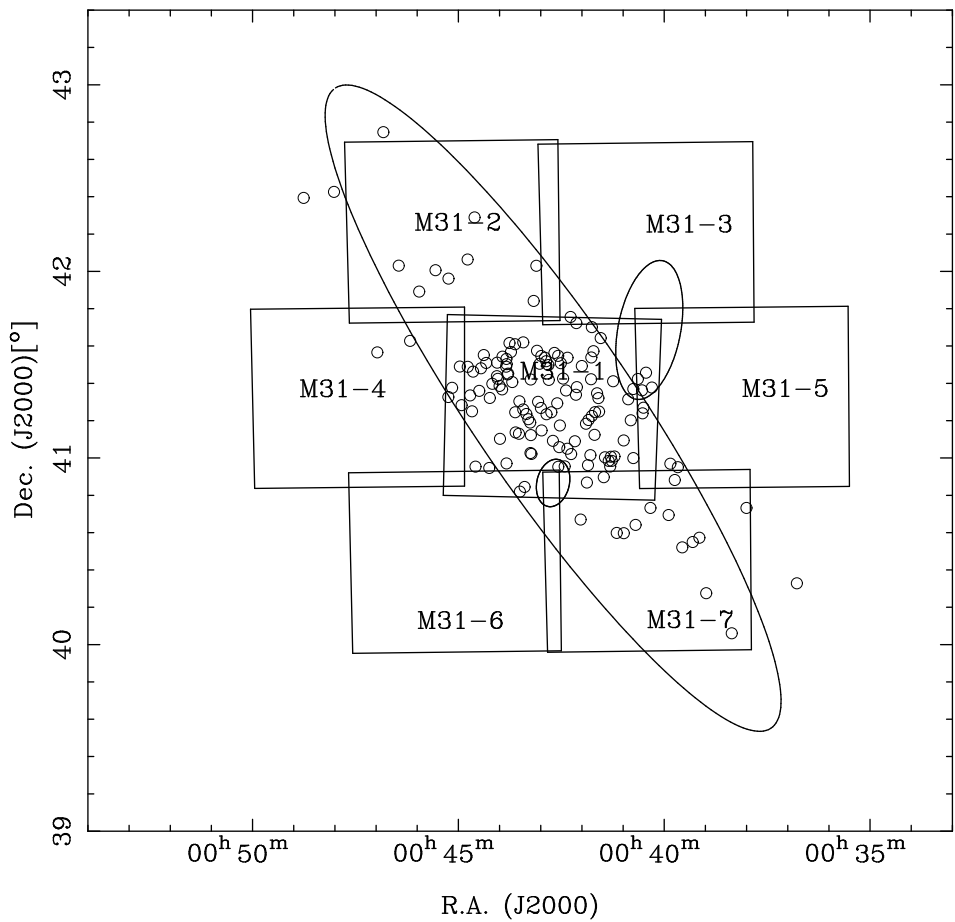}
\caption[]{The spatial distribution of our sample of M31 GCs
represented with small circles. The big ellipse marked the boundary
of M31 disk \citep{rac91} while the other two small ellipses are
$D_{25}$ of NGC 205 (northwest) and M32 (southeast), respectively.
The seven large boxes represent the seven observational fields of
BATC imaging survey. The angular size of each BATC imaging field is
$\sim 58'\times 58'$. North is up and east is to the left.}
\label{fig1}
\end{figure}

Because the spectroscopic metallicities of \citet{per02} were based
on the [Fe/H] calibration of \citet{bh90}, who only included old GCs
in their calibration sample, only the metallicities of the old GCs
in \citet{per02} are reliable. Further, it has long been known that
the degeneracy of age and metallicity for SEDs and color fitting of
the star clusters \citep[see e.g. ][]{wor94,jiang03,fan06,ma07}.
However, fortunately, the SED/colors are only weakly correlated with
age, if the GCs are old enough. Thus, if we only study the old GCs,
the SED/colors are only a function of metallicity, which could
eliminate effects of the other factors. For these reasons, we just
used the old confirmed GCs with observed metallcities to avoid these
problems and ensure our results reliable. This resulted in 144
matches between the the sample of 416 old confirmed clusters from
\citet{peac10} and the spectroscopic metallicity sample of
\citet{per02}. All the matched M31 GCs are shown in Fig. \ref{fig1}
with BATC observational field overlaid. The matched GCs are marked
with small circles, of which four are out of BATC field of view.

\begin{figure}
\center
\includegraphics[scale=.8,angle=-90]{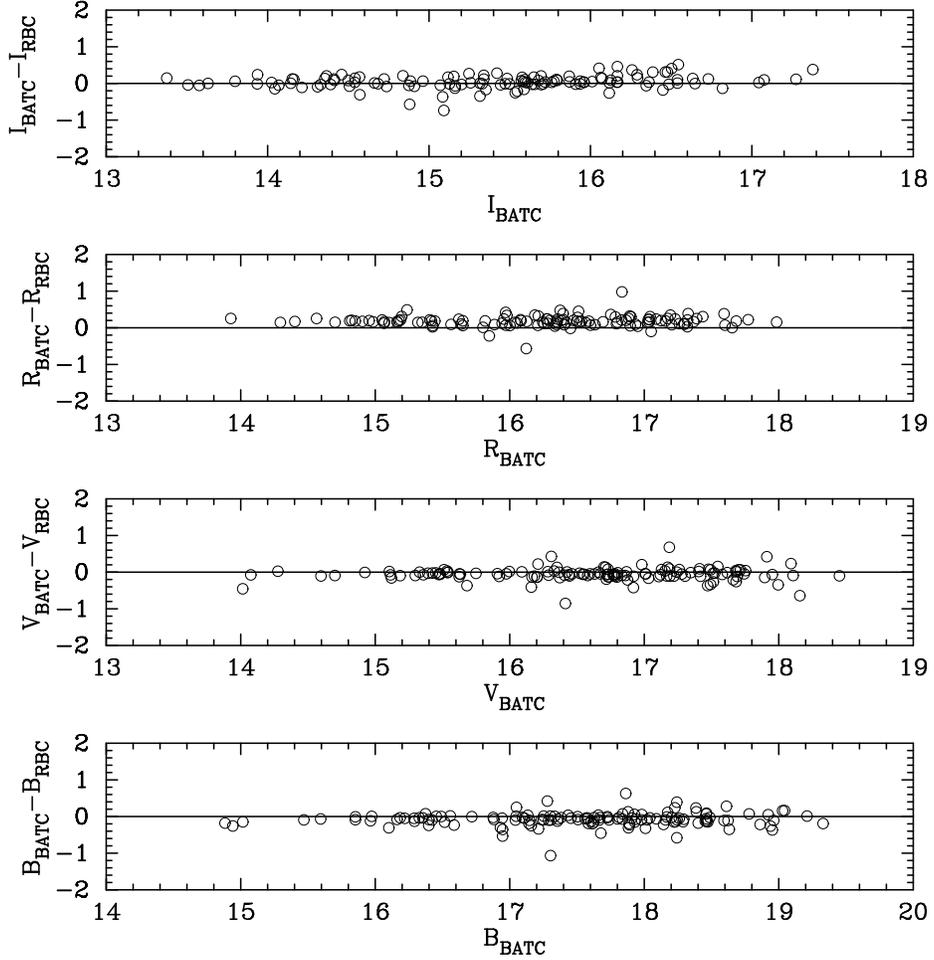}
\caption[]{The comparisons of broadband magnitudes from BATC and
those from RBC v.4.0. The mean magnitude offsets are
$\overline{B_{BATC}-B_{RBC}}=-0.08\pm0.20$,
$\overline{V_{BATC}-V_{RBC}}=-0.08\pm0.30$,
$\overline{R_{BATC}-R_{RBC}}=0.14\pm0.35$,
$\overline{I_{BATC}-I_{RBC}}=0.04\pm0.19$. The figure indicates that
the our photometry are basically consistent with the previous
photometry from RBC v.4.0.} \label{fig2}
\end{figure}

For the comparisons with the previous photometry of these M31 GCs,
we transformed the BATC intermediate-band system to the $UBVRI$
broad-band system using the transformation formulas between these
two systems from \citet{zhou03}. The formulas were derived based on
the BATC observations of the Landolt standard stars from the
catalogs of \citet{land83,land92} and \citet{ga00} in fifteen
colors. Since the photometry in the BATC $a$ and $b$ filters are not
used in our work, we cannot obtain the $U$ magnitudes from BATC
photometry. Fig. \ref{fig2} shows the comparisons of the $B$, $V$,
$R$, and $I$ photometry in our work with previous measurements from
RBC v.4.0. The mean magnitude offsets are
$\overline{B_{BATC}-B_{RBC}}=-0.08\pm0.20$,
$\overline{V_{BATC}-V_{RBC}}=-0.08\pm0.30$,
$\overline{R_{BATC}-R_{RBC}}=0.14\pm0.35$,
$\overline{I_{BATC}-I_{RBC}}=0.04\pm0.19$. From this figure it can
be seen that the $B$, $V$ and $I$ derived from BATC photometry are
consistent with that from RBC v.4.0, although there are some scatter
points. It also might be noticed that the offset of the $R$-band
photometry is slightly larger than that for the other band. This may
be due to the errors in the transformation formula by using the BATC
photometry of the Landolt standard stars, since the spectrum of a
globular cluster and a star is different. Further, we should be
aware of heterogeneous photometry summarized in RBC v.4.0 although
the authors made great effort to make it uniform, which might still
introduce the offset in the comparisons. In general, we believe that
our photometry is consistent with the previous studies and it will
not produce bias or affect our results in our work.

\subsection{The $ugriz$K Photometry from literature}
\label{sect:sam}

As we described in Sect. \ref{sect:intr}, \citet{peac10} identified
M31 GCs with the SDSS archive and images taken from Wide Field
CAMera (WFCAM) on the UK Infrared Telescope. The authors used the
SExtractor \citep{ba96} for the identifications and photometry of
sources in SDSS images and they found good consistency with the
previous photometry. The K-band images were reduced with the WFCAM
pipeline and the magnitudes were calibrated with the 2MASS Point
Source Catalogue. Thus, we draw the attention of readers that the
$ugriz$ photometry of \citet{peac10} is in AB-based photometric
system while for the K photometry is in Vega-based photometric
system. In our work, we did not transform all the magnitudes into
the same (AB-based or Vega-based) photometric system because such
color shift will not affect the correlations coefficients in our
analysis. Further this way could keep consistency and avoid
confusion. Finally the authors listed the photometry for 416 old
confirmed M31 GCs in the SDSS $ugriz$ bands and 2MASS K band, which
may be the latest work for study of M31 GCs. This photometry
catalogue will be utilized for our following analysis.

\subsection{The Reddenings of M31 GCs}
\label{sect:red}

In order to obtain the intrinsic colors for the sample GCs, all the
colors derived in our work should be corrected by reddening. Here we
adopted the reddening from \citet{barmby00} and \citet{fan08}, which
are the two most comprehensive reddening catalogue for M31 GCs so
far. \citet{barmby00} calibrated the relations of the intrinsic
color and metallicity and the $Q-$parameters for the Galactic GCs
and then applied the relations to M31. The authors assumed that the
extinction laws of M31 and Milky Way are the same, and they proved
that the assumption is reasonable, which indicates their results are
reliable. Finally, \citet{barmby00} determined the reddenings for
314 of M31 GCs and GC candidates, of which 221 are reliable (from
the private communication). Later, \citet{fan08} used the similar
method with the photometry of RBC v.3.5 \citep{gall04} and enlarged
the reddening sample of M31 GCs to 443 ones, half of which have
$E(B-V)\leq0.2$. Moreover, \citet{fan08} find good agreement between
their reddening with the previously determined reddening of
\citet{barmby00}. These two reddening catalogues of M31 GCs will be
applied for our following study.

\section{Analysis and Results}
\label{sect:ana}

In this section, we investigate the correlations between
spectroscopic metallicities and the BATC colors; further we analyzed
the similar correlations for $ugriz$K colors in \citet{peac10}. The
main goal of this paper is to find the most metal-sensitive and most
metal-nonsensitive colors, which might provide a method for roughly
estimating the metallicities just by colors.

\subsection{The Correlations of BATC colors and metallicities}
\label{sect:batcco}

From the 140 GCs in the BATC field of Fig. \ref{fig1}, there are
still nine clusters which we cannot obtain accurate photometric
measurements due to nearby bright objects (B065-G126, B091D-D058,
B160-G214 and B072), very low signal-to-noise (B305-D024), or steep
gradients in the galaxy light near the nucleus (B129). Moreover,
there are some problems with the photometry of three GCs: B127-G185,
B225-G280 and B306-G029. We found "emission lines" in their SEDs,
however, after careful checking of images, it turns out to be not
real. For these reasons, we exclude these nine clusters from our
sample and reduce the number of sample GCs to 131, which have the
reliable BATC photometry.

Because we only concern the intrinsic colors, all the observed
colors obtained in our work are required to be corrected by
reddening. As recalled in Sect. \ref{sect:red}, the reddening of M31
GCs from \citet{fan08} and \citet{barmby00} will be applied for the
estimation of intrinsic colors. For each object, we searched and
adopted the reddening from \citet{fan08} as priority, since it is
homogeneous and the number of GCs included is greater than that of
\citet{barmby00}. If a GC does not have the reddening in
\citet{fan08}, we used the reddening of \citet{barmby00} instead.
Considering the reddening can affect the color (hence the
correlations) significantly, the GCs which do not have reddenings
from either \citet{fan08} or \citet{barmby00} are excluded from our
analysis. After the elimination, we obtain 123 old confirmed GCs
with known reddening, which will be used as our final sample for the
analysis of correlations between the metallicity and BATC colors.

As discussed in Sect. \ref{sect:batc}, the images in the thirteen
BATC bands (from $c$ to $p$ band) were used in our work, thus
seventy-eight different colors could be derived and used in the
correlations analysis. We plotted the correlations between all the
BATC intrinsic color and the metallicities from \citet{per02} in
Fig.\ref{fig3}. We performed the linear regressions of the intrinsic
colors against metallicity correlations for the 123 GCs of M31 with
an equation below,
\begin{equation}
(x-y)_0=a+b{\rm[Fe/H]}
\end{equation}
\noindent Here we use $(x-y)_0$ as general notation to represent any
intrinsic BATC color. It is worth noting that there are several
outliers in Fig.\ref{fig3}, which seems not to follow the general
trend very well. This may be due to the errors on the metallicity,
inaccurate reddening estimates, or the errors on the photometry.
Further, although the GCs in our sample are old and the SEDs almost
the same, the slight difference of ages also might cause such
scatter.

Table \ref{t1.tab} lists the BATC intrinsic colors, intercede $a$
and slope $b$, the correlation coefficients $r$, fitting point
numbers of our sample of M31 GCs. Finally, we find that correlation
coefficients of 23 colors $r>0.7$. In particular, even for the
colors $(f-k)_0$, $(f-o)_0$, $(h-k)_0$, the correlation coefficients
$r>0.8$. These metal-sensitive colors could be used for estimating
the metallicity in the future work. However, it should also be noted
that the correlation coefficients approach zero for $(g-h)_0$,
$(k-m)_0$, $(k-n)_0$, $(m-n)_0$, which are nearly independent of
metallicity. This result can be explained by there being less
absorption lines in the corresponding wavelength span.

\begin{figure}
\center
\includegraphics[scale=.8,angle=0]{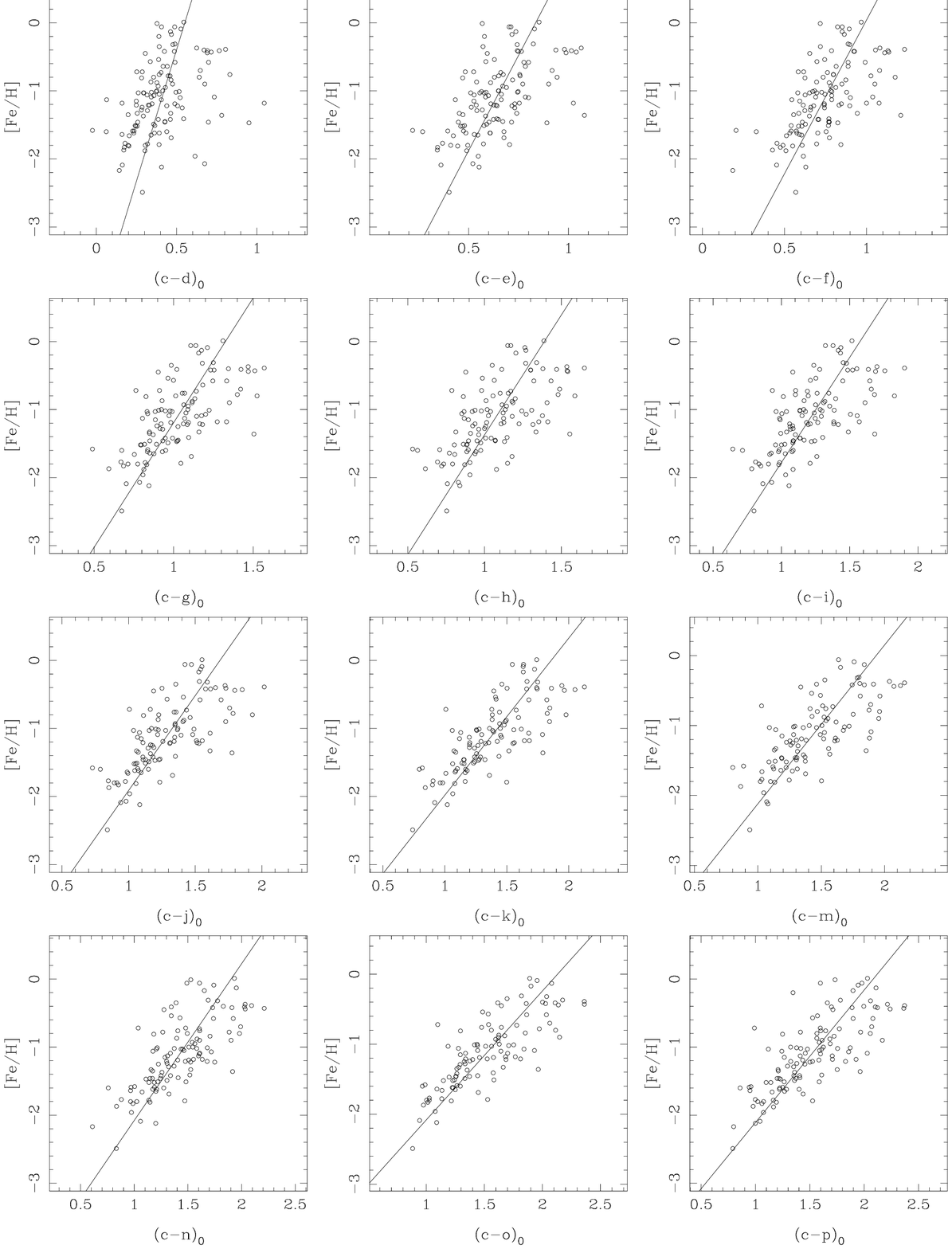}
\caption[]{The correlations of spectroscopic metallicities from
\citet{per02} and BATC intrinsic colors of M31 GCs.} \label{fig3}
\end{figure}
\addtocounter{figure}{-1}
\begin{figure}
\center
\includegraphics[scale=.8,angle=0]{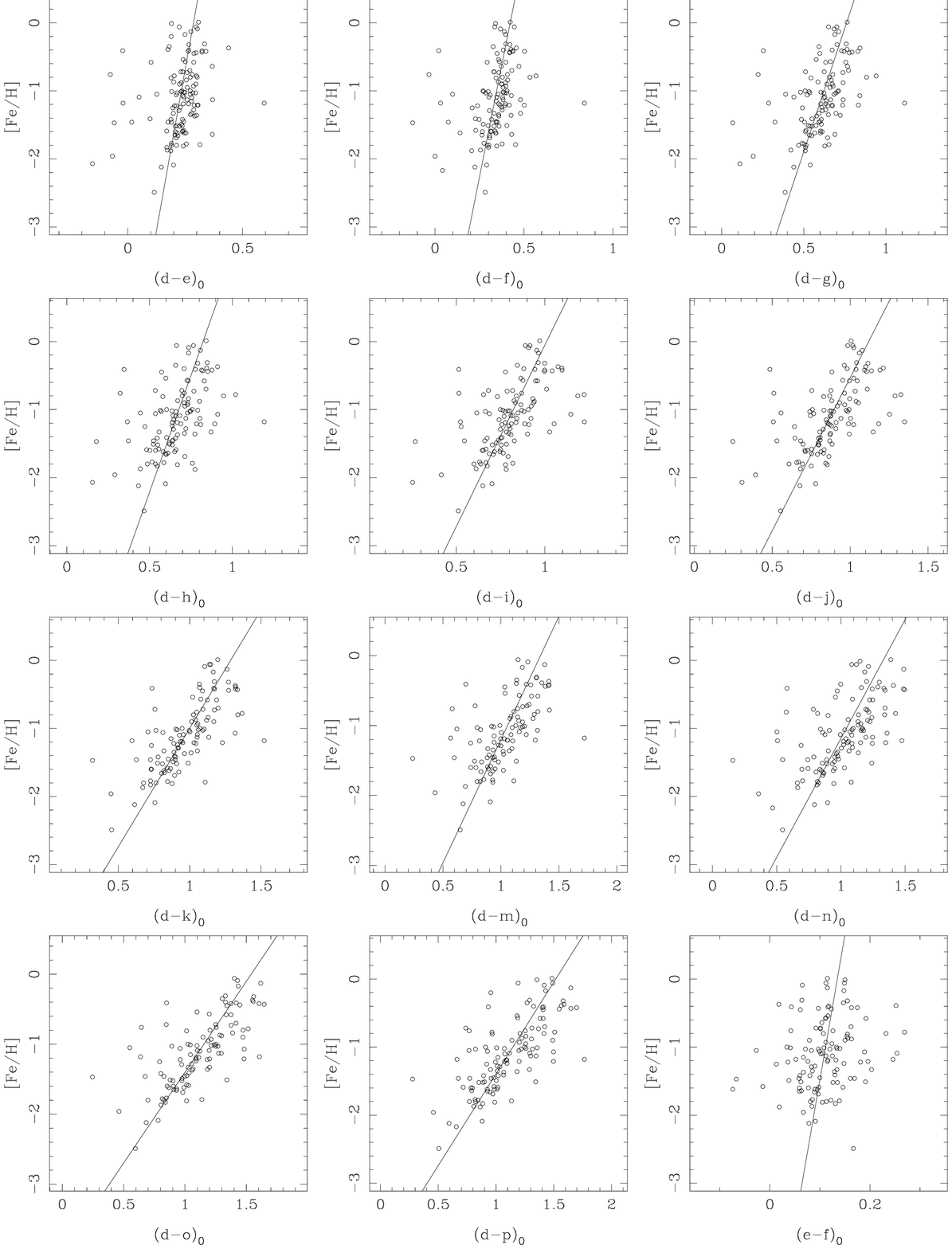}
\caption[]{Continued.}
\end{figure}
\addtocounter{figure}{-1}
\begin{figure}
\center
\includegraphics[scale=.8,angle=0]{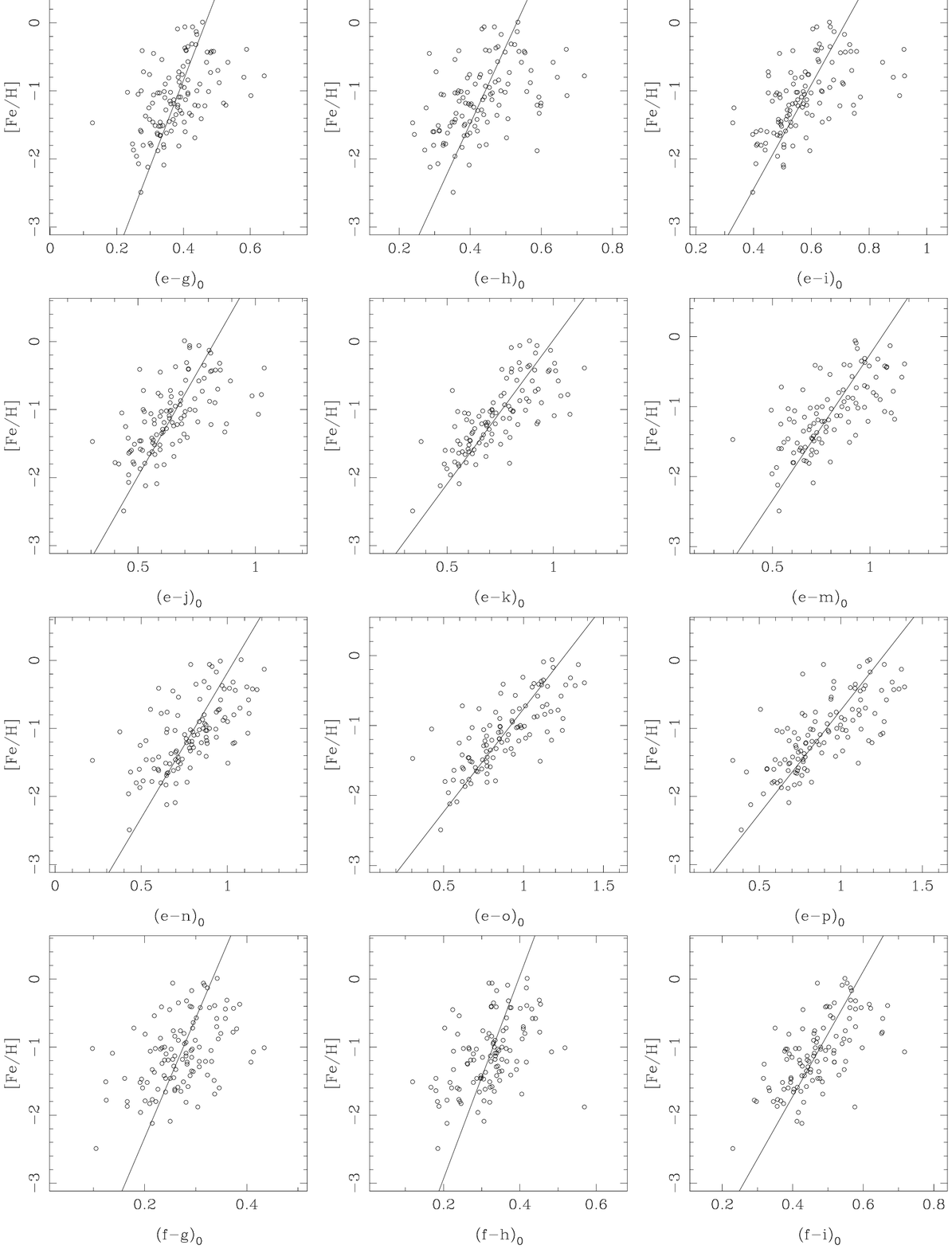}
\caption[]{Continued.}
\end{figure}
\addtocounter{figure}{-1}
\begin{figure}
\center
\includegraphics[scale=.8,angle=0]{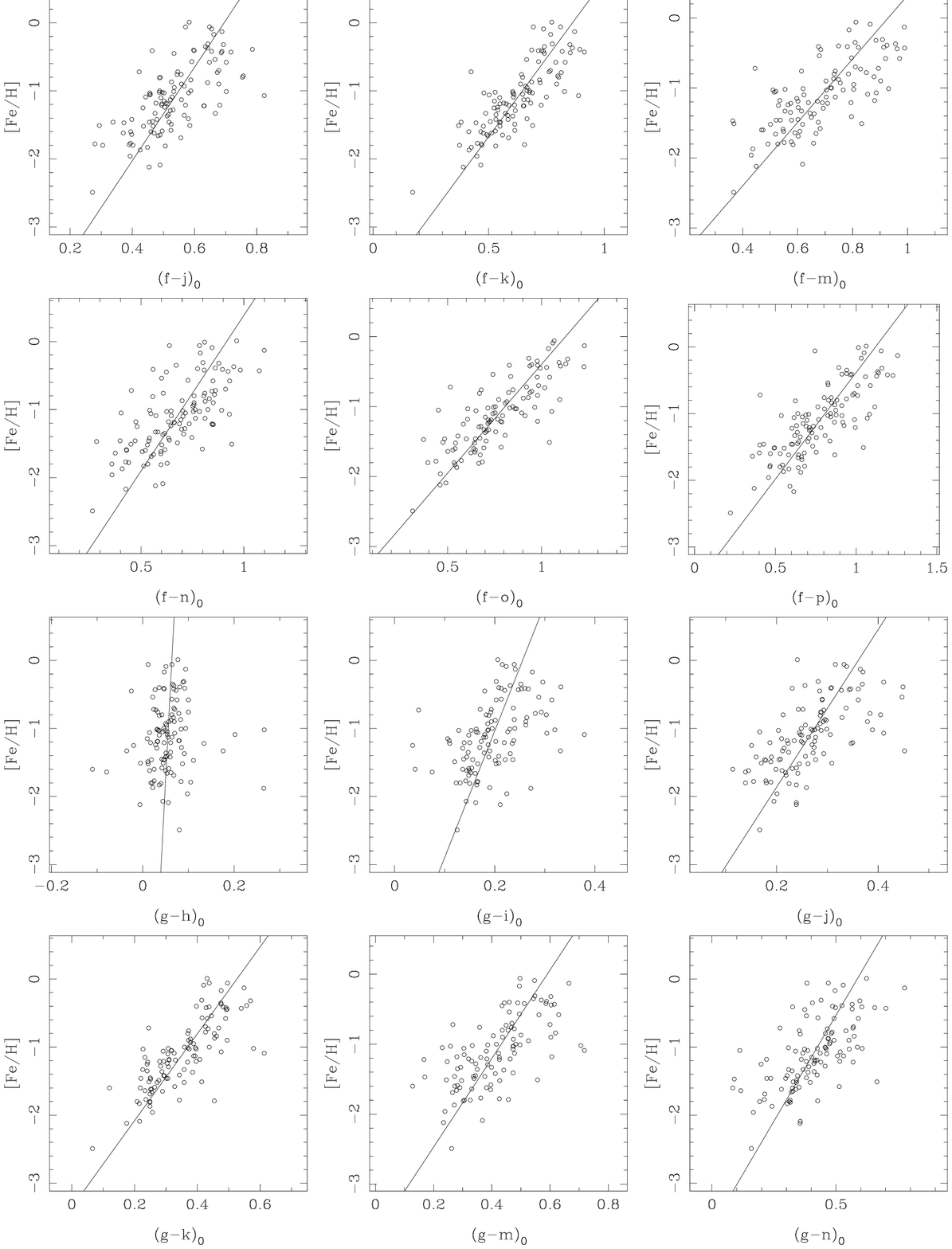}
\caption[]{Continued.}
\end{figure}
\addtocounter{figure}{-1}
\begin{figure}
\center
\includegraphics[scale=.8,angle=0]{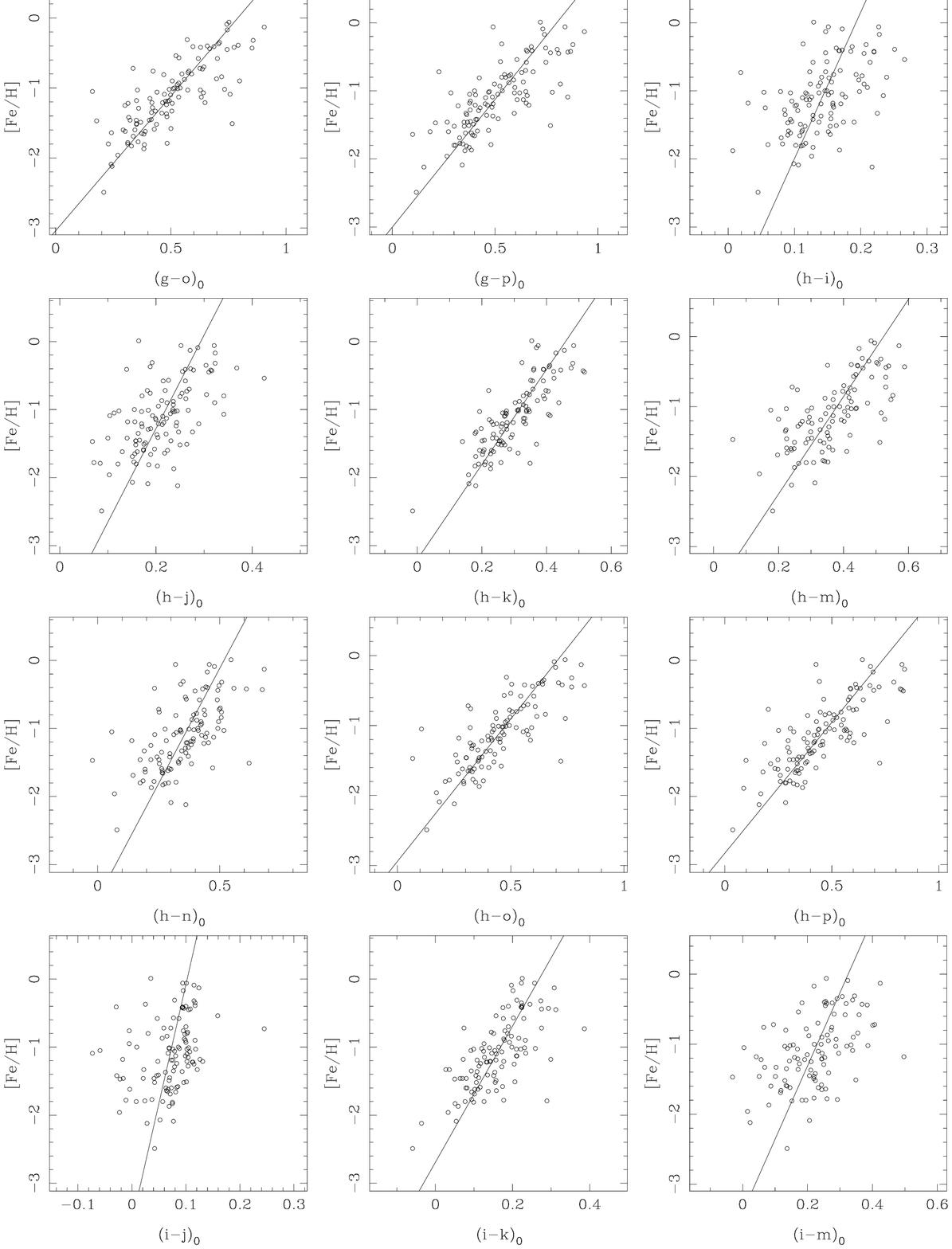}
\caption[]{Continued.}
\end{figure}
\addtocounter{figure}{-1}
\begin{figure}
\center
\includegraphics[scale=.8,angle=0]{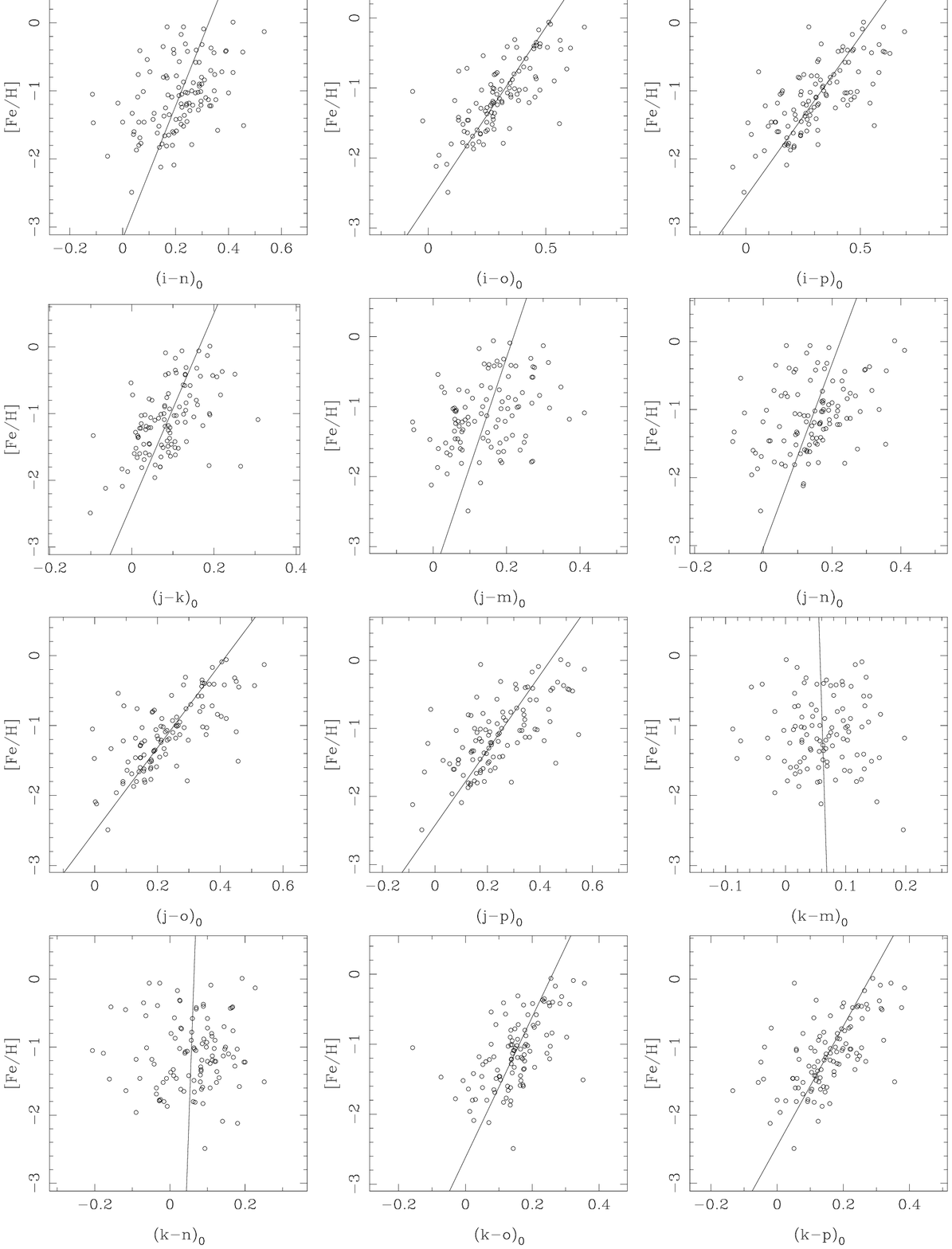}
\caption[]{Continued.}
\end{figure}
\addtocounter{figure}{-1}
\begin{figure}
\center
\includegraphics[scale=.8,angle=0]{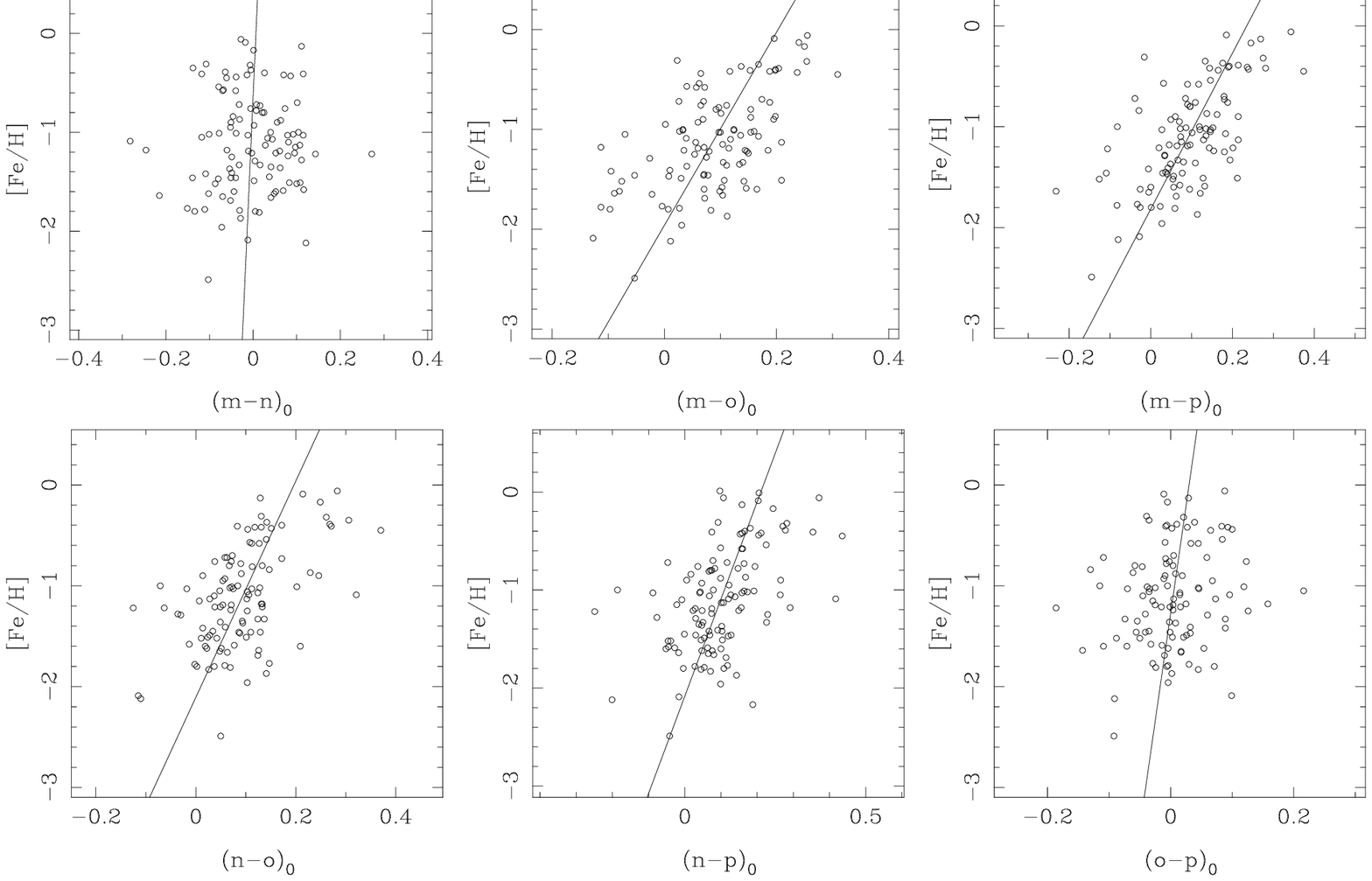}
\caption[]{Continued.}
\end{figure}

\begin{table}
\caption{The correlation coefficients of observational spectroscopic
metallicities from \citet{per02} and BATC intrinsic colors.
$(x-y)_0= a+b{\rm[Fe/H]}$.} \label{t1.tab}
\begin{center}
\begin{tabular}{ccccc}
\hline \hline
BATC colors &   \multicolumn{1}{c}{$a$} & \multicolumn{1}{c}{$b$} & $ r$ & $N$ \\
\hline
$(c-d)_0$ & $  0.551\pm 0.035$ & $  0.130\pm 0.029$ &    0.383&  121\\
$(c-e)_0$ & $  0.836\pm 0.030$ & $  0.179\pm 0.024$ &    0.563&  119\\
$(c-f)_0$ & $  0.991\pm 0.033$ & $  0.221\pm 0.027$ &    0.607&  118\\
$(c-g)_0$ & $  1.325\pm 0.037$ & $  0.273\pm 0.030$ &    0.653&  114\\
$(c-h)_0$ & $  1.390\pm 0.041$ & $  0.284\pm 0.033$ &    0.627&  113\\
$(c-i)_0$ & $  1.575\pm 0.042$ & $  0.323\pm 0.034$ &    0.668&  113\\
$(c-j)_0$ & $  1.686\pm 0.043$ & $  0.359\pm 0.035$ &    0.702&  111\\
$(c-k)_0$ & $  1.859\pm 0.044$ & $  0.433\pm 0.037$ &    0.754&  109\\
$(c-m)_0$ & $  1.934\pm 0.050$ & $  0.441\pm 0.041$ &    0.726&  104\\
$(c-n)_0$ & $  1.902\pm 0.047$ & $  0.434\pm 0.039$ &    0.724&  115\\
$(c-o)_0$ & $  2.131\pm 0.050$ & $  0.543\pm 0.041$ &    0.791&  107\\
$(c-p)_0$ & $  2.083\pm 0.050$ & $  0.514\pm 0.041$ &    0.759&  117\\
$(d-e)_0$ & $  0.286\pm 0.020$ & $  0.053\pm 0.016$ &    0.290&  120\\
$(d-f)_0$ & $  0.423\pm 0.024$ & $  0.076\pm 0.020$ &    0.333&  119\\
$(d-g)_0$ & $  0.761\pm 0.030$ & $  0.138\pm 0.025$ &    0.465&  115\\
$(d-h)_0$ & $  0.823\pm 0.031$ & $  0.146\pm 0.025$ &    0.478&  114\\
$(d-i)_0$ & $  1.010\pm 0.031$ & $  0.187\pm 0.025$ &    0.575&  114\\
$(d-j)_0$ & $  1.119\pm 0.036$ & $  0.223\pm 0.030$ &    0.584&  112\\
$(d-k)_0$ & $  1.286\pm 0.034$ & $  0.287\pm 0.028$ &    0.700&  110\\
$(d-m)_0$ & $  1.342\pm 0.045$ & $  0.285\pm 0.037$ &    0.601&  105\\
$(d-n)_0$ & $  1.328\pm 0.044$ & $  0.286\pm 0.036$ &    0.596&  116\\
$(d-o)_0$ & $  1.540\pm 0.047$ & $  0.385\pm 0.038$ &    0.699&  108\\
$(d-p)_0$ & $  1.517\pm 0.043$ & $  0.371\pm 0.035$ &    0.699&  118\\
$(e-f)_0$ & $  0.135\pm 0.012$ & $  0.023\pm 0.010$ &    0.212&  117\\
$(e-g)_0$ & $  0.466\pm 0.016$ & $  0.079\pm 0.013$ &    0.492&  115\\
$(e-h)_0$ & $  0.531\pm 0.019$ & $  0.088\pm 0.016$ &    0.466&  114\\
$(e-i)_0$ & $  0.719\pm 0.021$ & $  0.130\pm 0.017$ &    0.576&  114\\
$(e-j)_0$ & $  0.827\pm 0.025$ & $  0.165\pm 0.020$ &    0.616&  112\\
$(e-k)_0$ & $  0.995\pm 0.025$ & $  0.236\pm 0.020$ &    0.747&  110\\
$(e-m)_0$ & $  1.061\pm 0.032$ & $  0.240\pm 0.027$ &    0.665&  105\\
$(e-n)_0$ & $  1.043\pm 0.032$ & $  0.234\pm 0.027$ &    0.636&  115\\
$(e-o)_0$ & $  1.259\pm 0.035$ & $  0.340\pm 0.029$ &    0.756&  108\\
$(e-p)_0$ & $  1.240\pm 0.034$ & $  0.328\pm 0.028$ &    0.733&  117\\
\hline
\end{tabular}
\end{center}
\end{table}
\addtocounter{table}{-1}
\begin{table}
\caption{Continued.}
\begin{center}
\begin{tabular}{ccccc}
\hline \hline
BATC color &   \multicolumn{1}{c}{$a$} & \multicolumn{1}{c}{$b$} & $r$ & $N$ \\
\hline
$(f-g)_0$ & $  0.333\pm 0.013$ & $  0.057\pm 0.010$ &    0.458&  114\\
$(f-h)_0$ & $  0.397\pm 0.015$ & $  0.067\pm 0.013$ &    0.449&  113\\
$(f-i)_0$ & $  0.587\pm 0.015$ & $  0.109\pm 0.013$ &    0.634&  112\\
$(f-j)_0$ & $  0.695\pm 0.018$ & $  0.145\pm 0.015$ &    0.679&  111\\
$(f-k)_0$ & $  0.863\pm 0.019$ & $  0.217\pm 0.016$ &    0.800&  110\\
$(f-m)_0$ & $  0.929\pm 0.025$ & $  0.221\pm 0.021$ &    0.721&  105\\
$(f-n)_0$ & $  0.916\pm 0.028$ & $  0.219\pm 0.023$ &    0.671&  115\\
$(f-o)_0$ & $  1.128\pm 0.028$ & $  0.322\pm 0.023$ &    0.802&  108\\
$(f-p)_0$ & $  1.121\pm 0.030$ & $  0.313\pm 0.024$ &    0.768&  116\\
$(g-h)_0$ & $  0.064\pm 0.011$ & $  0.008\pm 0.009$ &    0.084&  113\\
$(g-i)_0$ & $  0.256\pm 0.012$ & $  0.054\pm 0.010$ &    0.458&  113\\
$(g-j)_0$ & $  0.361\pm 0.013$ & $  0.086\pm 0.010$ &    0.626&  112\\
$(g-k)_0$ & $  0.526\pm 0.015$ & $  0.157\pm 0.013$ &    0.767&  110\\
$(g-m)_0$ & $  0.591\pm 0.023$ & $  0.158\pm 0.019$ &    0.641&  105\\
$(g-n)_0$ & $  0.585\pm 0.023$ & $  0.161\pm 0.019$ &    0.619&  112\\
$(g-o)_0$ & $  0.790\pm 0.025$ & $  0.260\pm 0.020$ &    0.782&  108\\
$(g-p)_0$ & $  0.800\pm 0.026$ & $  0.266\pm 0.022$ &    0.761&  112\\
$(h-i)_0$ & $  0.193\pm 0.010$ & $  0.046\pm 0.008$ &    0.490&  112\\
$(h-j)_0$ & $  0.293\pm 0.012$ & $  0.073\pm 0.010$ &    0.575&  111\\
$(h-k)_0$ & $  0.459\pm 0.012$ & $  0.143\pm 0.010$ &    0.813&  109\\
$(h-m)_0$ & $  0.524\pm 0.019$ & $  0.144\pm 0.015$ &    0.677&  104\\
$(h-n)_0$ & $  0.517\pm 0.022$ & $  0.148\pm 0.018$ &    0.608&  112\\
$(h-o)_0$ & $  0.723\pm 0.023$ & $  0.246\pm 0.019$ &    0.782&  107\\
$(h-p)_0$ & $  0.737\pm 0.025$ & $  0.260\pm 0.020$ &    0.775&  113\\
$(i-j)_0$ & $  0.102\pm 0.010$ & $  0.029\pm 0.008$ &    0.322&  111\\
$(i-k)_0$ & $  0.269\pm 0.012$ & $  0.100\pm 0.010$ &    0.690&  109\\
$(i-m)_0$ & $  0.325\pm 0.021$ & $  0.096\pm 0.017$ &    0.481&  104\\
$(i-n)_0$ & $  0.325\pm 0.024$ & $  0.103\pm 0.020$ &    0.444&  112\\
$(i-o)_0$ & $  0.527\pm 0.023$ & $  0.199\pm 0.019$ &    0.714&  107\\
$(i-p)_0$ & $  0.542\pm 0.024$ & $  0.212\pm 0.020$ &    0.710&  112\\
$(j-k)_0$ & $  0.165\pm 0.014$ & $  0.070\pm 0.011$ &    0.516&  109\\
$(j-m)_0$ & $  0.219\pm 0.021$ & $  0.064\pm 0.017$ &    0.343&  105\\
$(j-n)_0$ & $  0.224\pm 0.021$ & $  0.074\pm 0.017$ &    0.381&  111\\
\hline
\end{tabular}
\end{center}
\end{table}
\addtocounter{table}{-1}
\begin{table}
\caption{Continued.}
\begin{center}
\begin{tabular}{ccccc}
\hline \hline
BATC color &   \multicolumn{1}{c}{$a$} & \multicolumn{1}{c}{$b$} & $r$ & $N$ \\
\hline
$(j-o)_0$ & $  0.420\pm 0.020$ & $  0.167\pm 0.016$ &    0.706&  108\\
$(j-p)_0$ & $  0.439\pm 0.023$ & $  0.181\pm 0.019$ &    0.673&  110\\
$(k-m)_0$ & $  0.057\pm 0.013$ & $ -0.004\pm 0.011$ &   $-0.033$&  104\\
$(k-n)_0$ & $  0.064\pm 0.020$ & $  0.006\pm 0.017$ &    0.036&  110\\
$(k-o)_0$ & $  0.261\pm 0.017$ & $  0.100\pm 0.014$ &    0.580&  106\\
$(k-p)_0$ & $  0.279\pm 0.018$ & $  0.114\pm 0.015$ &    0.600&  109\\
$(m-n)_0$ & $  0.006\pm 0.021$ & $  0.010\pm 0.017$ &    0.058&  105\\
$(m-o)_0$ & $  0.203\pm 0.018$ & $  0.103\pm 0.015$ &    0.573&  105\\
$(m-p)_0$ & $  0.235\pm 0.019$ & $  0.129\pm 0.016$ &    0.634&  104\\
$(n-o)_0$ & $  0.196\pm 0.017$ & $  0.093\pm 0.014$ &    0.534&  108\\
$(n-p)_0$ & $  0.210\pm 0.021$ & $  0.101\pm 0.017$ &    0.479&  115\\
$(o-p)_0$ & $  0.030\pm 0.015$ & $  0.023\pm 0.012$ &    0.186&  107\\
\hline
\end{tabular}
\end{center}
\end{table}

\subsection{The Correlations of $ugriz$K colors and metallicities}

Similar to the procedure in Sect. \ref{sect:batcco}, in the
following section, we will analyze the correlations between SDSS
$ugriz$ and K colors from \citet{peac10} and the spectroscopic
metallicities of \citet{per02}. In total, we obtained 127 matches
between the old confirmed GCs sample in Fig. \ref{fig1} and the
reddening values of Sect. \ref{sect:red}, which is the final sample
for the fittings of the correlations between metallicities and
$ugrizK$ colors. In order to thoroughly study the correlations with
these photometry, we investigated all the fifteen different colors,
for each of which, we plot the linear fitting between the color and
metallicities with Eq.(1) in Fig. \ref{fig4}. Similar to Table
\ref{t1.tab}, the intercede $a$, slope $b$ and the correlation
coefficients $r$ are summarized in Table \ref{t2.tab}. Apparently,
all the correlation coefficients (\emph{r}) are high ($r>0.7$),
suggesting the $ugriz$K colors are very metal-sensitive. The lowest
correlation coefficient is $r_{(i-z)}=0.701$ while the highest one
is $r_{(u-\rm {K})}=0.845$, which might be due to $(u-$K$)_0$ is the
widest wavelength coverage color and most absorption lines are
included in this range. Whilst, for the $(i-z)_0$ color, the
absorption line number should be the least.

It is interesting that \citet{peng06} studied the correlation of
[Fe/H] and the $(g-z)_0$ for 95 GCs from Milky Way, M87 and M49, and
they found that the correlation is very likely nonlinear. Instead,
they used a broken linear relation to fit their data. However, such
fitting also suffers from scatter and the constraint is not very
robust. In fact, we cannot tell which fitting (either the linear or
the nonlinear) is significantly better than the other from Fig.
\ref{fig4}. In this case, our purpose is to find the metal-sensitive
colors and thus the one-order approximation is sufficient.

\begin{figure}
\center
\includegraphics[scale=.8,angle=0]{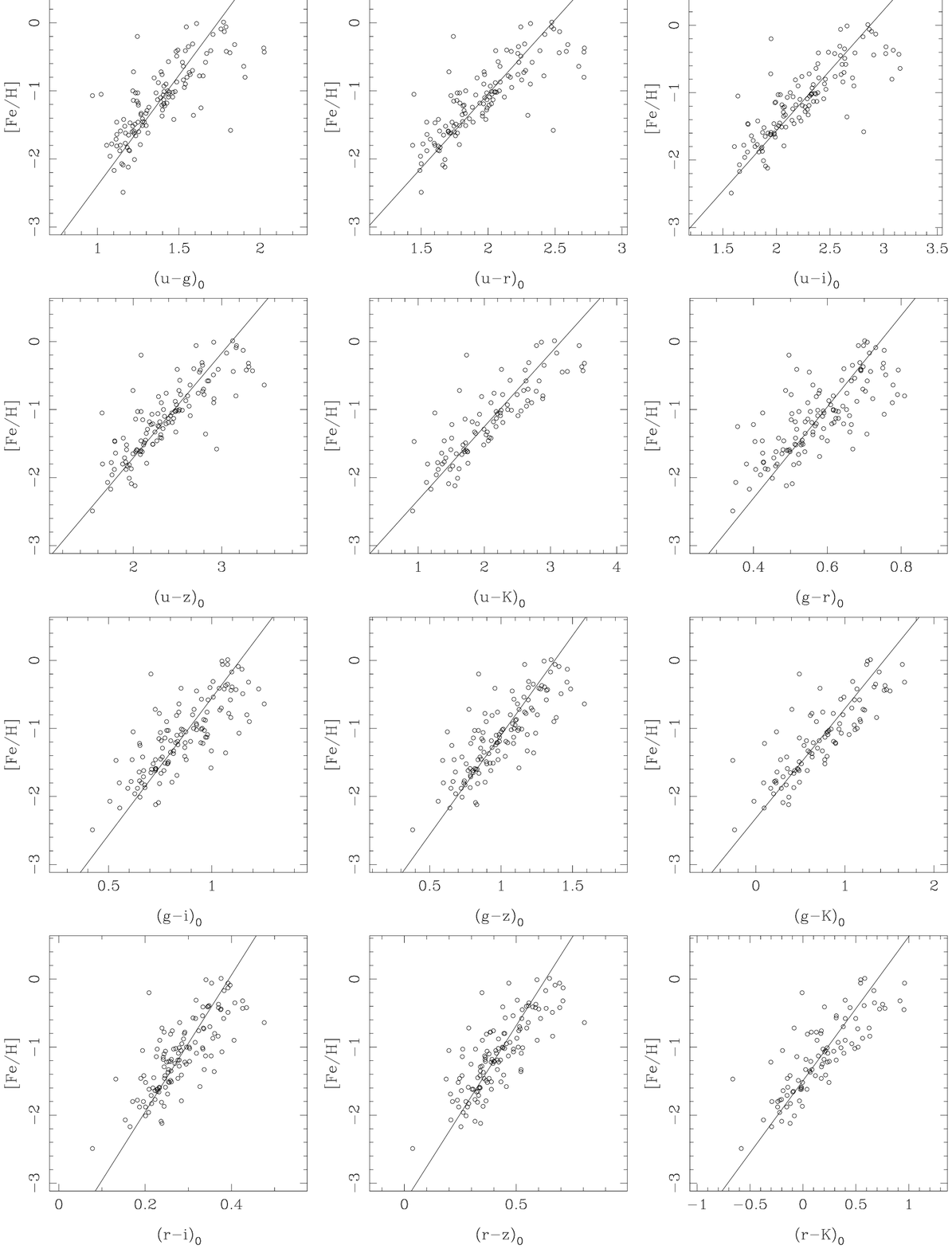}
\caption[]{The correlations of spectroscopic metallicities from
\citet{per02} and intrinsic SDSS $ugriz$ and K colors from
\citet{peac10}.} \label{fig4}
\end{figure}
\addtocounter{figure}{-1}
\begin{figure}
\center
\includegraphics[scale=.8,angle=0]{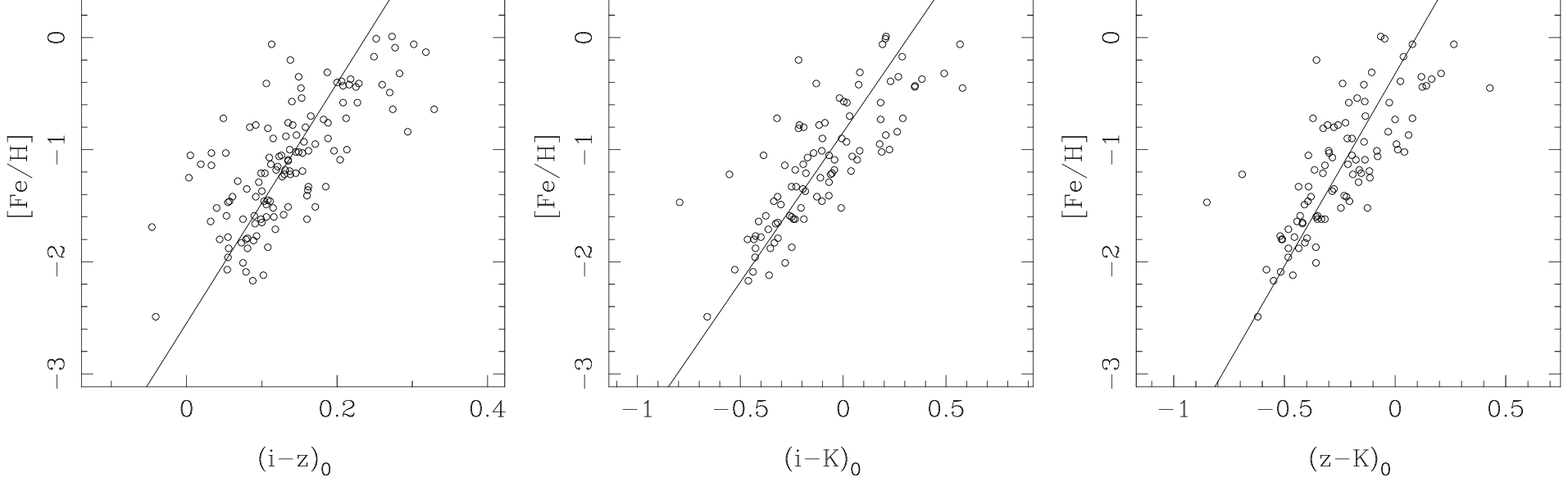}
\caption[]{Continued.}
\end{figure}

\begin{table}
\caption{The correlation coefficients of the observational
spectroscopic metallicities from \citet{per02} and SDSS $ugriz$ and
K colors from \citet{peac10}. $(x-y)_0= a+b{\rm[Fe/H]}$.}
\label{t2.tab}
\begin{center}
\begin{tabular}{ccccc}
\hline \hline
$ugriz$K colors &   \multicolumn{1}{c}{$a$} & \multicolumn{1}{c}{$b$} & $ r$ & $N$ \\
\hline
$(g-i)_0$ & $  1.137\pm 0.022$ & $  0.248\pm 0.018$ &    0.775&  127\\
$(g-\rm K)_0$ & $  1.439\pm 0.055$ & $  0.620\pm 0.043$ &    0.828&   96\\
$(g-r)_0$ & $  0.743\pm 0.015$ & $  0.149\pm 0.012$ &    0.740&  127\\
$(g-z)_0$ & $  1.375\pm 0.029$ & $  0.342\pm 0.023$ &    0.796&  127\\
$(i-\rm K)_0$ & $  0.316\pm 0.039$ & $  0.374\pm 0.030$ &    0.786&   96\\
$(i-z)_0$ & $  0.238\pm 0.011$ & $  0.093\pm 0.008$ &    0.701&  127\\
$(r-i)_0$ & $  0.394\pm 0.009$ & $  0.099\pm 0.007$ &    0.770&  127\\
$(r-\rm K)_0$ & $  0.705\pm 0.045$ & $  0.472\pm 0.035$ &    0.808&   96\\
$(r-z)_0$ & $  0.632\pm 0.017$ & $  0.193\pm 0.014$ &    0.785&  127\\
$(u-g)_0$ & $  1.739\pm 0.031$ & $  0.309\pm 0.025$ &    0.753&  122\\
$(u-i)_0$ & $  2.874\pm 0.046$ & $  0.559\pm 0.037$ &    0.807&  122\\
$(u-\rm K)_0$ & $  3.160\pm 0.080$ & $  0.927\pm 0.062$ &    0.845&   92\\
$(u-r)_0$ & $  2.480\pm 0.039$ & $  0.459\pm 0.032$ &    0.797&  122\\
$(u-z)_0$ & $  3.112\pm 0.052$ & $  0.651\pm 0.042$ &    0.816&  122\\
$(z-\rm K)_0$ & $  0.094\pm 0.034$ & $  0.292\pm 0.027$ &    0.750&   96\\
\hline
\end{tabular}
\end{center}
\end{table}

\section{Summary}
\label{sect:sum}

Our work provides the most metal-sensitive colors of M31 GCs which
could be utilized for roughly estimating the metallicity of GCs from
their colors. The observational spectroscopic metallicities applied
in our analysis were from \citet{per02} and the colors are derived
from BATC photometry and \citet{peac10}.

First, we performed the photometry with BATC archival images from
$c$ to $p$ band, covering the wavelength from $\sim$ 4,000 to $\sim$
10,000 {\AA} and obtained the BATC colors for 123 confirmed old GCs.
The reddening values for the color corrections are from
\citet{fan08} and \citet{barmby00}. For the thorough study, all the
seventy-eight different BATC colors are used for the correlation
analysis. Our fitting result shows that correlation coefficients of
23 colors $r>0.7$. Especially, for the colors $(f-k)_0$, $(f-o)_0$,
$(h-k)_0$, the correlation coefficients $r>0.8$. Meanwhile, the
correlation coefficients approach zero for $(g-h)_0$, $(k-m)_0$,
$(k-n)_0$, $(m-n)_0$, which can be explained by lacking absorption
lines.

Further, we also fitted the correlations of metallicity and $ugriz$K
colors for the 127 old confirmed GCs with the photometry from
\citet{peac10}. Similarly, all the fifteen colors are utilized for
analysis. The fitting result indicates that all these colors are
metal-sensitive with correlation coefficients $r>0.7$. In
particular, $(u-$K$)_0$ is the most metal-sensitive color while
$(i-z)_0$ is the metal-nonsensitive color.

\acknowledgments We are very grateful to an anonymous Referee for
her/his useful comments and suggestions. This research was supported
by the Chinese National Natural Science Foundation through Grant
Nos. 10873016, 10803007, 10778720, 10633020, 10673012 and by
National Basic Research Program of China (973 Program) No.
2007CB815403. Z.F. thanks the Young Researcher Grant of National
Astronomical Observatories, Chinese Academy of Sciences.

\end{document}